\documentclass{sig-alternate-05-2015}

\usepackage{tabularx}
\usepackage{xcolor}
\usepackage{subfigure}

\usepackage{url} 

\newcommand{\Fig}[1]{Fig.~\ref{fig:#1}}
\newcommand{\Sec}[1]{Sec.~\ref{sec:#1}}
\newcommand{\Tab}[1]{Tab.~\ref{tab:#1}}

\let\tempone\itemize
\let\temptwo\enditemize
\renewenvironment{itemize}{\tempone\addtolength{\itemsep}{-0.7\baselineskip}}{\temptwo}
\let\tempthree\enumerate
\let\tempfour\endenumerate
\renewenvironment{enumerate}{\tempthree\addtolength{\itemsep}{-0.7\baselineskip}}{\tempfour}

\begin{document}

\title{
How Close to the Edge? Delay/Utilization Trends in MEC
}

\numberofauthors{3} 
\author{
\alignauthor
Francesco Malandrino\\
       \affaddr{Politecnico di Torino, Italy}
\alignauthor
Scott Kirkpatrick\\
       \affaddr{The Hebrew University of Jerusalem, Israel}
\alignauthor
Carla-Fabiana Chiasserini 
\\
       \affaddr{Politecnico di Torino, Italy\\
       CNR-IEIIT, Torino, Italy}
} 

\CopyrightYear{2016}
\setcopyright{acmcopyright}
\conferenceinfo{CAN'16,}{December 12 2016, Irvine, CA, USA}
\isbn{978-1-4503-4673-3/16/12}\acmPrice{\$15.00}
\doi{http://dx.doi.org/10.1145/3010079.3010080}
\clubpenalty=10000 
\widowpenalty = 10000 

\maketitle

\begin{abstract}
Virtually all of the rapidly increasing data traffic consumed by mobile users requires some kind of processing, normally performed at cloud servers.  A recent thrust, {\em mobile edge computing}, moves such processing to servers {\em within} the cellular mobile network.  The large temporal and spatial variations to which mobile data usage is subject could make the reduced latency that edge clouds offer come at an unacceptable cost in redundant and underutilized infrastructure.  We present some first empirical results on this question, based on large scale sampled crowd-sourced traces from several major cities spanning multiple operators and identifying the applications in use.  We find opportunities to obtain both high server utilization and low application latency, but the best approaches will depend on the individual network operator's deployment strategy and geographic specifics of the cities we study.
\end{abstract}
\keywords{Mobile Edge Computing; Real-World Traces}

\section{Introduction}

All Internet traffic is generated through one or (typically) more processing steps. Videos, as an example, are transcoded to fit the device that will play them; web pages are added customized advertisement and location-specific content; social networking pages are essentially assembled on-the-fly based on the user's identity and location.

Traditionally, all this processing was performed by the individual content providers (e.g., YouTube) within their own premises. More recently, the bulk of the processing has been moved to Internet-based clouds.
Mobile edge computing (MEC) is an emerging paradigm that envisions moving cloud services (i.e., the servers) as close to mobile users as possible. Under the MEC approach, cloud servers would not be hosted at centralized datacenters but rather within the mobile network itself, from backhaul nodes to base stations~\cite{noi-pof}. MEC improves service latency (intuitively, it takes a shorter time to reach a nearby server) and reduces the load on core networks (less data is transferred to/from the Internet).

One potential drawback of MEC is {\em server utilization}: intuitively, if we deploy many small servers, they are more likely to be underutilized than fewer, larger ones.  Servers covering wider areas can exploit the fact that traffic demand at different locations is not synchronized in time, will support a larger number of users, and thus attain a higher average utilization.

Unlike other technologies, adopting MEC is not a binary, all-or-nothing decision; instead, network operators can decide {\em how much} MEC they would like. No MEC at all is the present-day cloud computing scenario. Pushing MEC to the extreme could mean deploying servers at each cellular base station, literally as close to users as they can be. For applications engaged with uniquely local or very popular information, this would give the best performance, but its efficiency for the wider range of mobile applications is questionable.  Intermediate options are possible, feasible, and indeed likely to work better than either extreme.

Information on the actual traffic demand and its variations coming from mobile operators is difficult to obtain, limited, and typically subject to non-disclosure agreements~\cite{orange-d4d,Malandrino-TCS}. In this paper, we use a set of {\em crowd-sourced} traces, coming from the users of the We-Fi app~\cite{wefi}. Our traces include multiple mobile networks, multiple network technologies, and multiple cities. Furthermore, they report mobile traffic on a per-application basis.

In summary, the main contributions of this paper are:
\begin{itemize}
\item the usage of real-world, crowd-sourced traces to study the relationship between {\em the extent} to which the MEC paradigm is embraced and the resulting average server utilization and latency;
\item proposing a methodology to combine deployment and demand information to obtain latency and server utilization estimates;
\item presenting numeric results, spanning multiple cities, mobile operators, and applications.
\end{itemize}

We begin by reviewing related work in \Sec{relwork}. Then, we present our traces in \Sec{traces} and describe our methodology in \Sec{steps}. After presenting our numeric results in \Sec{results}, we conclude the paper in \Sec{conclusion}.

\section{Related work}
\label{sec:relwork}

Our study is connected to three main categories of prior work: works presenting real-world mobile traces and datasets; works studying mobile edge computing; works doing the latter using the first.

Many real-world traces come from volunteers, e.g., the MIT Reality Project~\cite{mit-reality} and the Nokia Mobile Challenge~\cite{nokia}. These traces include a great deal of valuable information; their main shortcoming is the limited number of participants (in the case of the Nokia Mobile Challenge, around two hundred). This scale is adequate to study, for example, user mobility or encounter patterns; however, studying a whole cellular network requires information about many more users. 
Mobile operators are typically reluctant to release demand and deployment information to the scientific community. An exception is represented by the Data For Development dataset by Orange~\cite{orange-d4d}, including mobility information for 50,000 users in Ivory Coast, as well as CDR (call-detail record) information for phone calls and SMS messages. However, the Orange trace only includes voice calls and SMS, and is severely restricted by heavy anonymization -- each ID encountered gets a new coded identity for each ``ego site'' to which they are a neighbor. In other cases~\cite{Malandrino-TCS} mobile operators do release demand or deployment information to individual research teams under non-disclosure agreements; however, these traces typically include only one operator and/or only one city.

Mobile-edge computing has been recently introduced~\cite{fog} as a way to move ``the cloud'', i.e., the servers processing mobile traffic, closer to users, thus reducing the latency and load of networks. Network Function Virtualization is widely regarded to as an enabling technology for MEC (see, e.g.,\cite{etsi-wp}). Recent works have studied the radio techniques needed to enable MEC~\cite{mec-radio}, its relationship to the Internet-of-things~\cite{mec-iot} and context-aware, next-generation networks~\cite{mec-5g}. Closer to our scenario, the authors of~\cite{moving} study how caches and servers should be placed in the network as its load changes over time. 
With regard to MEC and caching, a prominent application is mobile video streaming. As an example,~\cite{multiop-caching,cdn1} account for layered video coding techniques, and address the problem of placing the right layers at the right cache -- with~\cite{multiop-caching} also accounting for cooperation between operators. Other works~\cite{proactive-caching,wons} aim at {\em foreseeing} the content demand, in order to proactively populate caches or serve users.

Not many works exist that combine real-world traces and mobile edge computing. Among the most recent ones, \cite{noi-pof}~studies the price (in terms of additional infrastructure) of deploying caches within the cellular core network. Compared to this study, \cite{noi-pof} only focuses on caching and vehicular traffic,  and it only considers the dataset for the city of Los Angeles.

\section{Input data}
\label{sec:traces}

WeFi~\cite{wefi} is an Android app providing its users with information on nearby Wi-Fi access points, including their speed, encryption type, and level of reliability. In order to obtain such a service, WeFi users agree to disclose information about their location and activity, including the connections available to their mobile phones and the active apps.

These data represent a valuable snapshot of how networks and their users behave in the real world. In our paper, we use three datasets, relative to the American cities of Atlanta, Los Angeles, and San Francisco. Their main features are summarized in \Tab{datasets}.

\begin{table}[t!]
\caption{
Our datasets.
\label{tab:datasets}
} 
\begin{tabularx}{1\columnwidth}{|c|X|X|X|}
\hline
& Atlanta & Los Angeles & San Francisco \\
\hline\hline
Time of collection & Oct. 2014 & Oct. 2014 & Mar. 2015\\
\hline
Covered area [km$^2$] & $55\times 66$ & $46\times 73$ & $14\times 11$\\
\hline
Total traffic [TB] & 9.34 & 35.61 & 9.18\\
\hline
Number of records & 13 million & 81 million & 60 million\\
\hline
Unique users & 9,203 & 64,386 & 14,018\\
\hline
Unique cells & 12,615 & 36,09 & 14,728\\
\hline
\end{tabularx}
\end{table}

For each user, a new record is generated every time that one of the following happens:
(i) a one-hour period expires;
(ii) the user location changes;
(iii) the user switches between apps;
(iv) the smartphone switches between networks.

Each record of each dataset contains information on (i) time; (ii) (anonymized) user identity; (iii) GPS position; (iv) mobile operator, current cell ID and location area code (LAC); (v) app considered; (vi) amount of uploaded and downloaded data.

These traces are especially suitable to the kind of analysis we need to perform, for three main reasons:
\begin{itemize}
\item they come from multiple cities, and therefore multiple traffic profiles and network topologies;
\item they include multiple network operators, with different deployment strategies and types of infrastructure (e.g., micro- vs. macro-cells);
\item they contain information on the app generating each traffic flow.
\end{itemize}
We are therefore able to obtain results that are both {\em realistic}, as they come from real-world traces, and {\em general}, as they are obtained for a wide variety of geographic locations, mobile operators, and apps. These objectives are typically conflicting, e.g., theoretic models typically lack realism while using a single trace can yield scarcely general results.

\section{Methodology}
\label{sec:steps}

\subsection{Proxy metrics}

Our high-level goal is to characterize the evolution of
application latency and server utilization in next-generation, MEC-based cellular networks. In order to
perform a direct estimation,
we would need a very detailed knowledge of the networks themselves, including the core network topology, the type of servers that mobile operators can deploy, and the kind of processing each type of traffic requires. Regrettably, this information is unavailable, for two obvious reasons.  First, mobile operators do not disclose their core topologies.  Second, our objective is to understand how in-network data processing will affect {\em future} applications in next-generation networks, not present-day ones, so we will have to extrapolate and look for trends, without worrying about the fine details of present systems.

The most popular way of working around the first of these issues is guesswork, relying on best-practice core topologies~\cite{softcell,noi-pof} and randomly assigning in-network processing steps to traffic flows~\cite{softcell}.

We opt for a different approach, and identify two {\em proxy} metrics that we can extract directly from our datasets, without the need for further assumptions.  As a proxy for latency, we consider the {\em distance} between the base stations and the servers processing their traffic, since delays will depend most strongly on the number of network steps over which the data must pass. Similarly we characterize the effectiveness of server deployment and utilization through an {\em efficiency} metric, defined as ratio between the {\em average and peak} traffic processed by said servers. We extract these metrics through a four-step process, as detailed in \Sec{processing-steps}.

It is important to point out that traffic itself is, in a way, used as a proxy metric. Indeed, we discuss the dimensioning and utilization of servers based on the amount of traffic (in bytes) they have to serve; however, capabilities of real-world servers are defined as a combination of CPU power, RAM, and storage capacity. Translating amounts of traffic into CPU and RAM requirements would require a precise knowledge of the processing needed by each type of traffic.  At that level of detail, one should also consider supporting different types of applications at different depths in the network, depending on the degree to which the data required is local and needed only close to a single base station, widely popular and used throughout the region, or individual and used both locally and infrequently.  These sorts of issues lie outside the scope of our study. Finally, although 3G and LTE traffic are distinguished in our traces, we have combined their traffic for this analysis.

\begin{figure*}[h!]
\centering
\subfigure[\label{fig:n_clusters}]{
    \includegraphics[width=.3\textwidth]{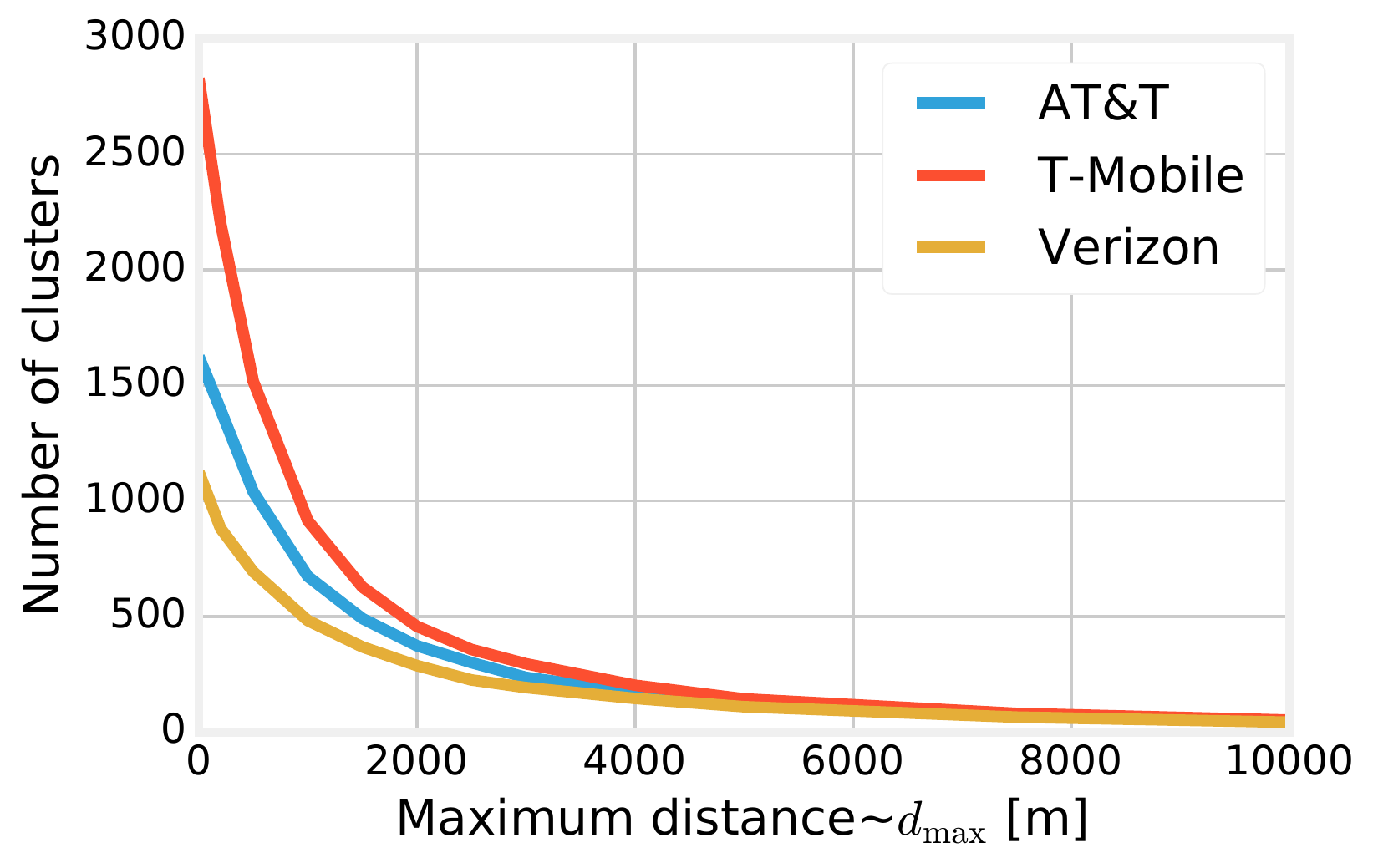}
}
\subfigure[\label{fig:bs_per_cluster}]{
    \includegraphics[width=.3\textwidth]{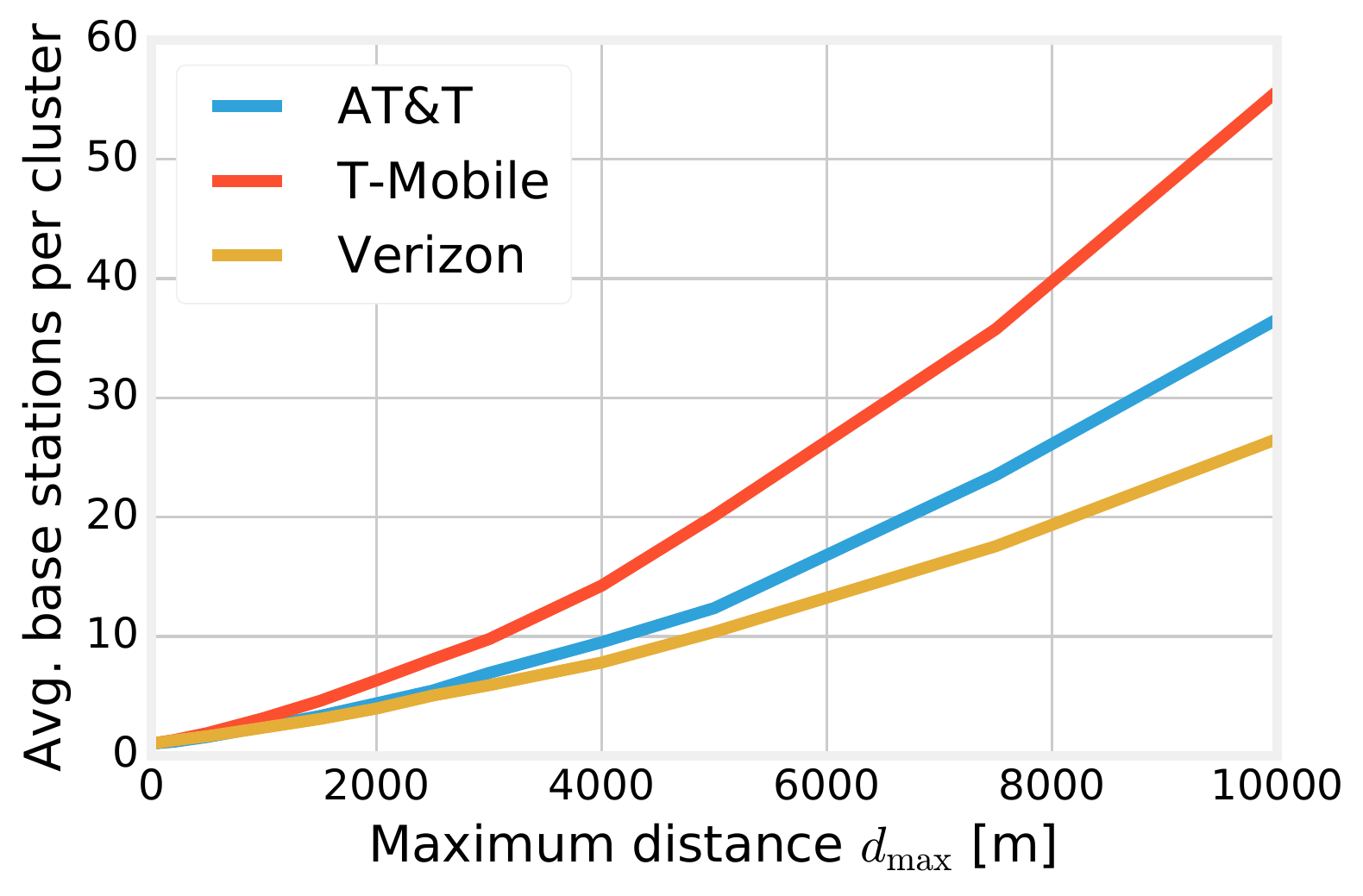}
}
\subfigure[\label{fig:map}]{
    \includegraphics[width=.3\textwidth]{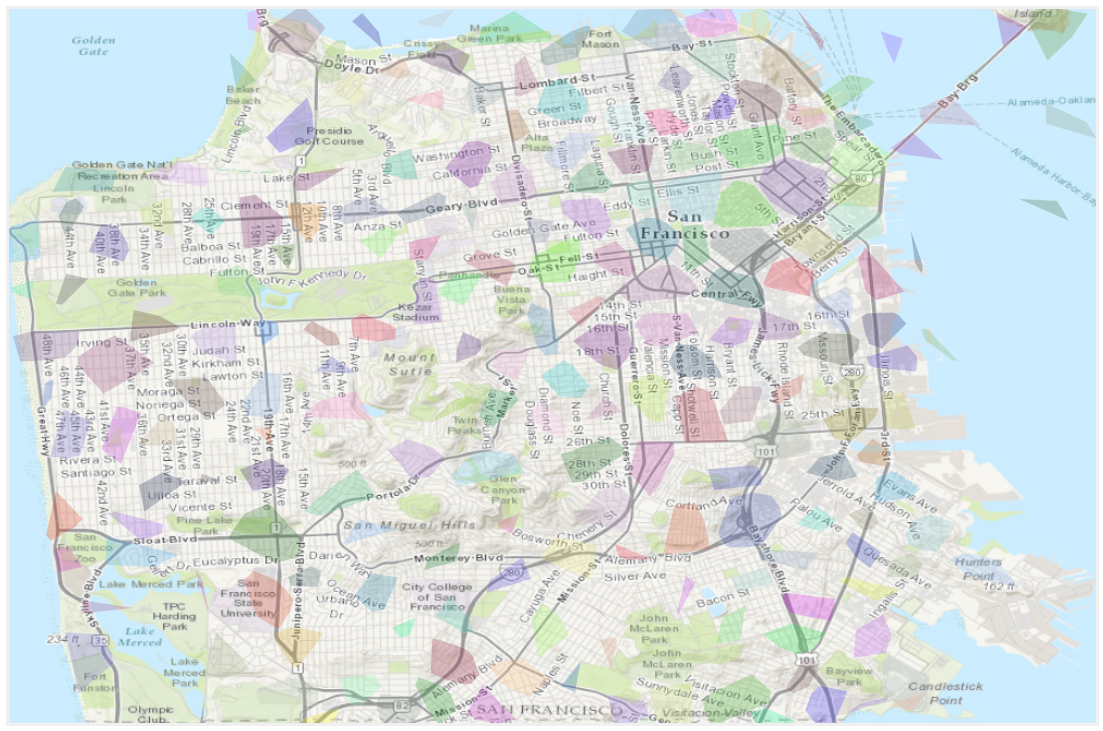}
}
\caption{
Clustering in San Francisco: number of clusters created as a function of the maximum distance~$d_{\max}$ (a); average number of base stations per cluster (b); location of the clusters created for T-Mobile when~$d_{\,ax}=1000~\text{m}$ (c). For clarity, in (c) we show the convex hull of the base stations in each cluster, not the coverage areas.
\label{fig:clustering}
} 
\end{figure*}

\begin{figure*}[h!]
\centering
\subfigure[\label{fig:eff-atl}]{
    \includegraphics[width=.3\textwidth]{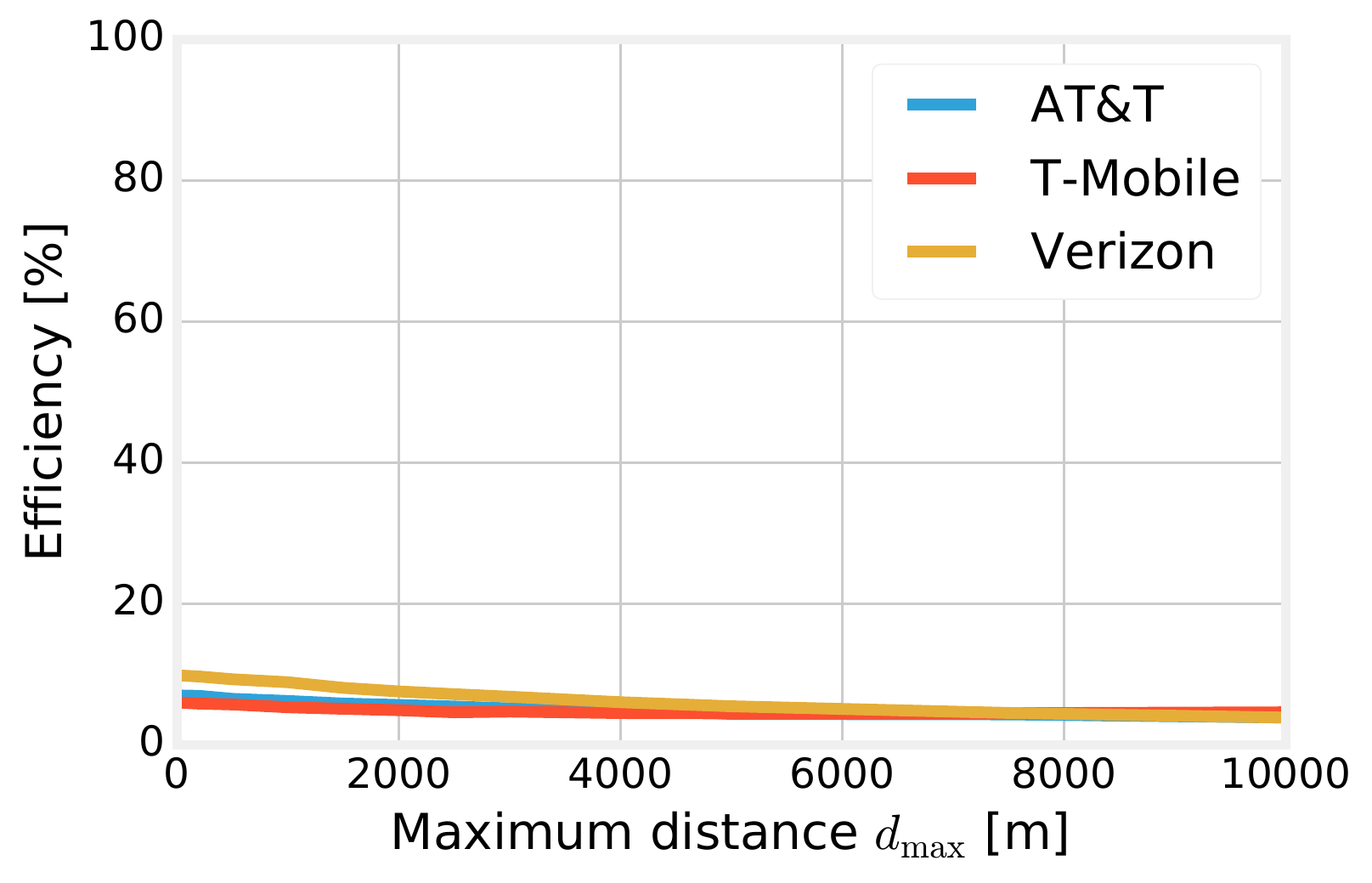}
}
\subfigure[\label{fig:eff-la}]{
    \includegraphics[width=.3\textwidth]{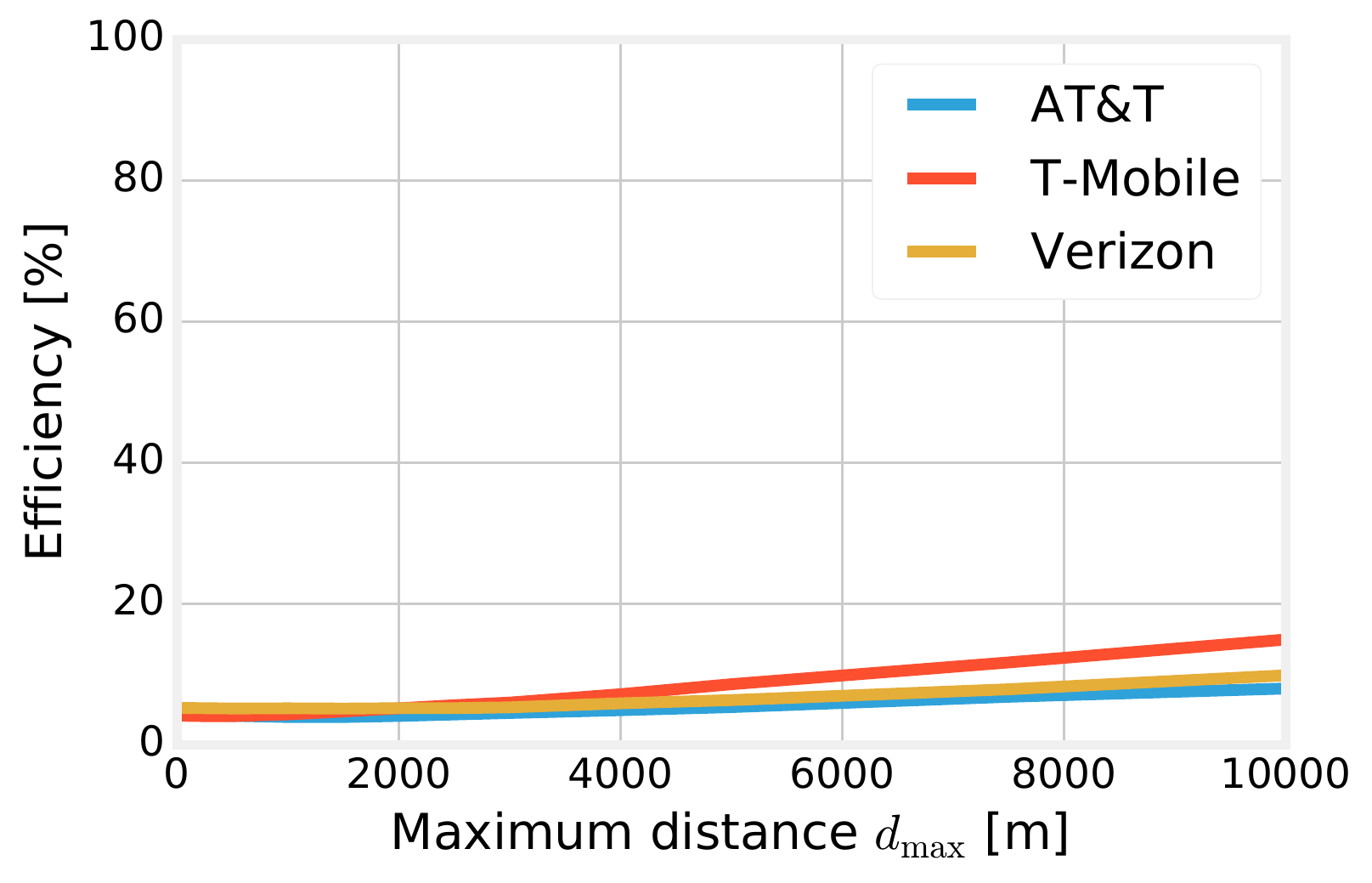}
}
\subfigure[\label{fig:eff-sf}]{
    \includegraphics[width=.3\textwidth]{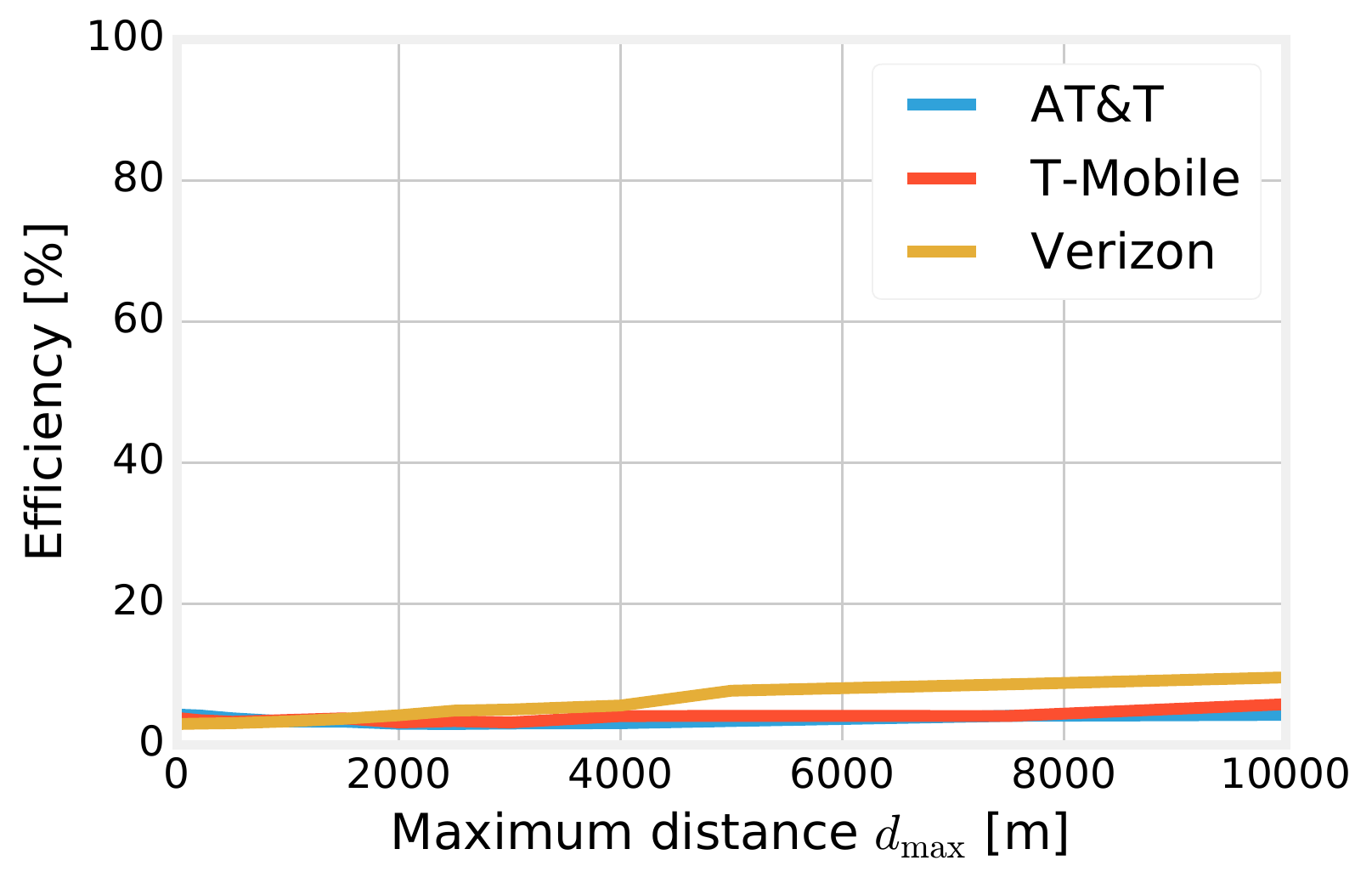}
}
\caption{
Efficiency as a function of~$d_{\max}$ for Atlanta (a); Los Angeles (b); San Francisco (c).
\label{fig:efficiency}
} 
\end{figure*}

\subsection{Processing steps}
\label{sec:processing-steps}

\paragraph*{Base stations}

Base stations are the interface between mobile networks and their users; therefore, knowing their position is of fundamental importance. Since this information is not directly included in our traces, we need to reconstruct it. To this end, we leverage two of the fields included in each record: the ID of the cell serving each user, and the user location.

More precisely, we associate to each cell a coverage area, corresponding to the convex hull of the locations of all users reporting being served by the cell itself. The base station serving each cell is located at the average of the coordinates of all users served by the cell itself, {\em weighted} by the amount of traffic they consume. This has the high-level effect of minimizing the distance between users and the base stations serving them.

\paragraph*{Traffic}

The next step deals with the traffic generated by users of each base station, for each day and hour. This information is readily available from our trace, so no special processing is needed. In addition to the total figures, we also keep track of the traffic coming from especially significant applications, identified from their clients running on mobile phones:
Facebook (client \path{COM.FACEBOOK.KATANA}),
YouTube (client \path{COM.GOOGLE.ANDROID.YOUTUBE}), and
Google Maps (client \path{COM.GOOGLE.ANDROID.APPS.MAPS}).

\paragraph*{Clustering}

As we mentioned, embracing the MEC paradigm is a continuous, rather than binary, decision. At one extreme of the spectrum, we can deploy one server for each base station; at the other, one server would serve the whole area under study. Adopting intermediate solutions corresponds to adjusting the set of base stations whose traffic is processed by each server.

We map this process to clustering, where each cluster corresponds to a server processing the traffic from one or more base stations. Specifically, we use a {\em hierarchical, bottom-up} clustering strategy known as the Voorhees algorithm~\cite{clustering},
as implemented in the well-known SciPy library (\path{http://scipy.org}).

The algorithm begins by assigning to each point its own {\em singleton} cluster. At each subsequent iteration:
\begin{enumerate}
\item for each pair of clusters, we compute an {\em inter-cluster distance}, defined as the maximum over all pairs of points such that the first point belongs to the first cluster and the second point to the second cluster;
\item if the minimum inter-cluster distance exceeds a threshold~$d_{\max}$, the algorithm terminates;
\item otherwise, the two clusters with the minimum inter-site distance are merged together, and we move to the next iteration.
\end{enumerate}

The resulting clustering corresponds to our MEC network design. Each cluster is no wider than~$d_{\max}$, is supported by one server, and the points belonging to that cluster are the base stations whose traffic is processed at the server. The parameter~$d_{\max}$ that we provide to the Voorhees algorithm is the first of our proxy variables: shorter distances correspond to lower latency, and vice versa.
Extreme values of~$d_{\max}$ correspond to the two extreme architectures we discussed earlier. When~$d_{\max}=0$, all traffic processing is performed at the individual base stations, thus having the most extreme possible MEC scenario. Very large values of $d_{\max}$, instead, correspond to scenarios where processing is highly centralized and there is no MEC at all.

\paragraph*{Efficiency}

The second proxy variable is the {\em efficiency} we observe with a certain clustering. This value is obtained as follows:
\begin{enumerate}
\item for each server, we consider the total traffic of the base stations it serves;
\item the maximum of such traffic over time represents the peak load the server must be dimensioned for;
\item the average of such traffic corresponds to the average load of the server;
\item the efficiency is the ratio of average to peak load.
\end{enumerate}

Our efficiency proxy metric would coincide with the actual server utilization if the servers' computational capabilities were dimensioned in such a way to match  their peak load. In practical cases server capabilities exceed their peak load, so our efficiency metric is slightly optimistic.
100\% efficiency is attained when the average load coincides with the peak load, i.e., when a server is always utilized at its maximum capacity. Servers that are underutilized for most of the time, instead, have efficiency values close to~0.

\subsection{Discussion}

Our methodology relies on proxy metrics, and this has the obvious disadvantage that our conclusions cannot be used to {\em quantitatively} study the impact of MEC. On the positive side, the {\em qualitative} conclusions we draw are not tainted by assumptions that cannot be verified or by unavoidable, educated guesses. 
Furthermore, our proxy metrics are strongly correlated to the original objectives of our study. While there are many factors affecting latency, the distance (in terms of both space and network nodes) between clients and servers is a major one. Similarly, the primary means of improving server utilization is to maximize the ratio between their average and peak load, i.e., our efficiency proxy metric.

Also notice that in all our steps we consider the networks of each mobile operator separately, as this allows us to ascertain whether their deployment and demand are so different that they react in different ways to MEC. Notice however that several sources~\cite{Malandrino-TCS,multiop-caching} expect that mobile operators (and content providers) will cooperate with each other in order to further improve their networks.

\section{Numeric results}
\label{sec:results}

We start by getting some additional insights on the clustering process and the role of its parameters, in \Sec{results-clustering}. Then, \Sec{results-efficiency} presents the relationship between the maximum distance~$d_{\max}$ and our efficiency proxy metric across different cities and mobile operators. Finally, in \Sec{results-applications} we go into further detail, by looking at traffic coming from different applications.

\subsection{The clustering process}
\label{sec:results-clustering}

\Fig{clustering} provides us with some insight on how the clustering process above works, and especially on the role of the $d_{\max}$ parameter\footnote{
  The plots in \Fig{clustering} are obtained for the San Francisco trace; the other traces under study yielded similar results.
}. At first sight, \Fig{n_clusters} and \Fig{bs_per_cluster} show nothing surprising: as the maximum distance increases, i.e., clusters can become bigger, we tend to create fewer of them (\Fig{n_clusters}) and clusters tend to include more base stations (\Fig{bs_per_cluster}). It is, however, worth noticing the three lines in each of the two plots, referring to three different mobile operators. 
What we  observe are clear, major differences between operators, sometimes a factor of three or four. This is due to the fact that, although all operators serve the whole  area under study, they do so through substantially different deployment strategies. For instance, Verizon has fewer base stations than other operators, each covering a wider area.

This further confirms the intuition that the effect of moving towards the MEC paradigm will be different for different mobile operators, and any decision in this respect should be made on a case-by-case basis. It also stresses the importance of real-world traces as a tool to study the operation and evolution of mobile networks -- especially if, as in our case, the traces include information for multiple mobile operators.

Finally, \Fig{map} reminds us of an important feature of our methodology. Clustering decisions are made solely on the grounds of the {\em position} of base stations, not their traffic. This corresponds to a real-world situation where servers are deployed based on connectivity considerations, and their features (e.g., memory and CPU) are dimensioned taking into account the traffic they will have to process.
More complex clustering algorithms, accounting for the actual traffic demand and the resulting load on servers, will be considered as a part of future work.

\begin{figure*}[h!]
\centering
\subfigure[\label{fig:example2}]{
    \includegraphics[width=.3\textwidth]{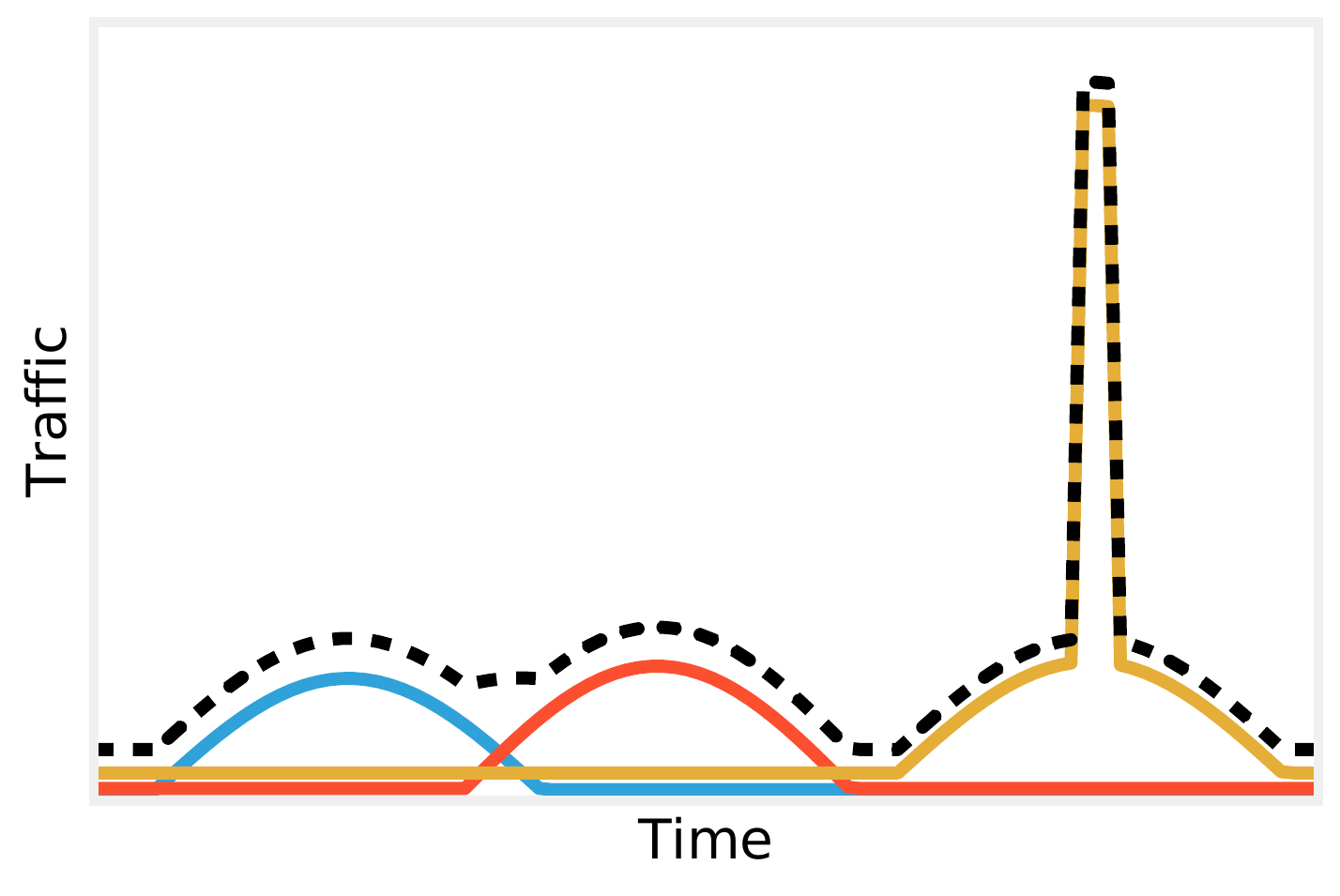}
}
\subfigure[\label{fig:loadcdf}]{
    \includegraphics[width=.3\textwidth]{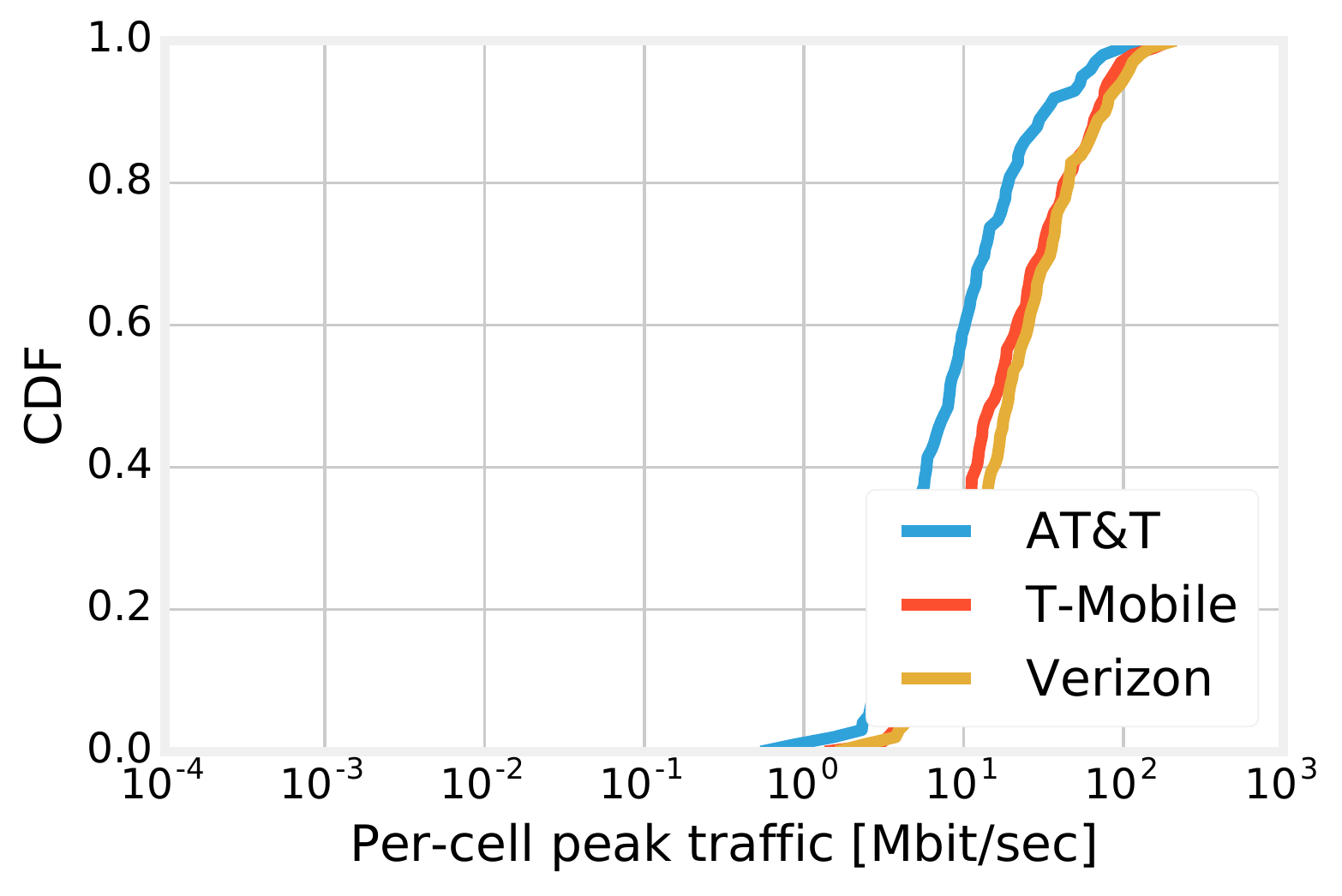}
}
\subfigure[\label{fig:ratios}]{
    \includegraphics[width=.3\textwidth]{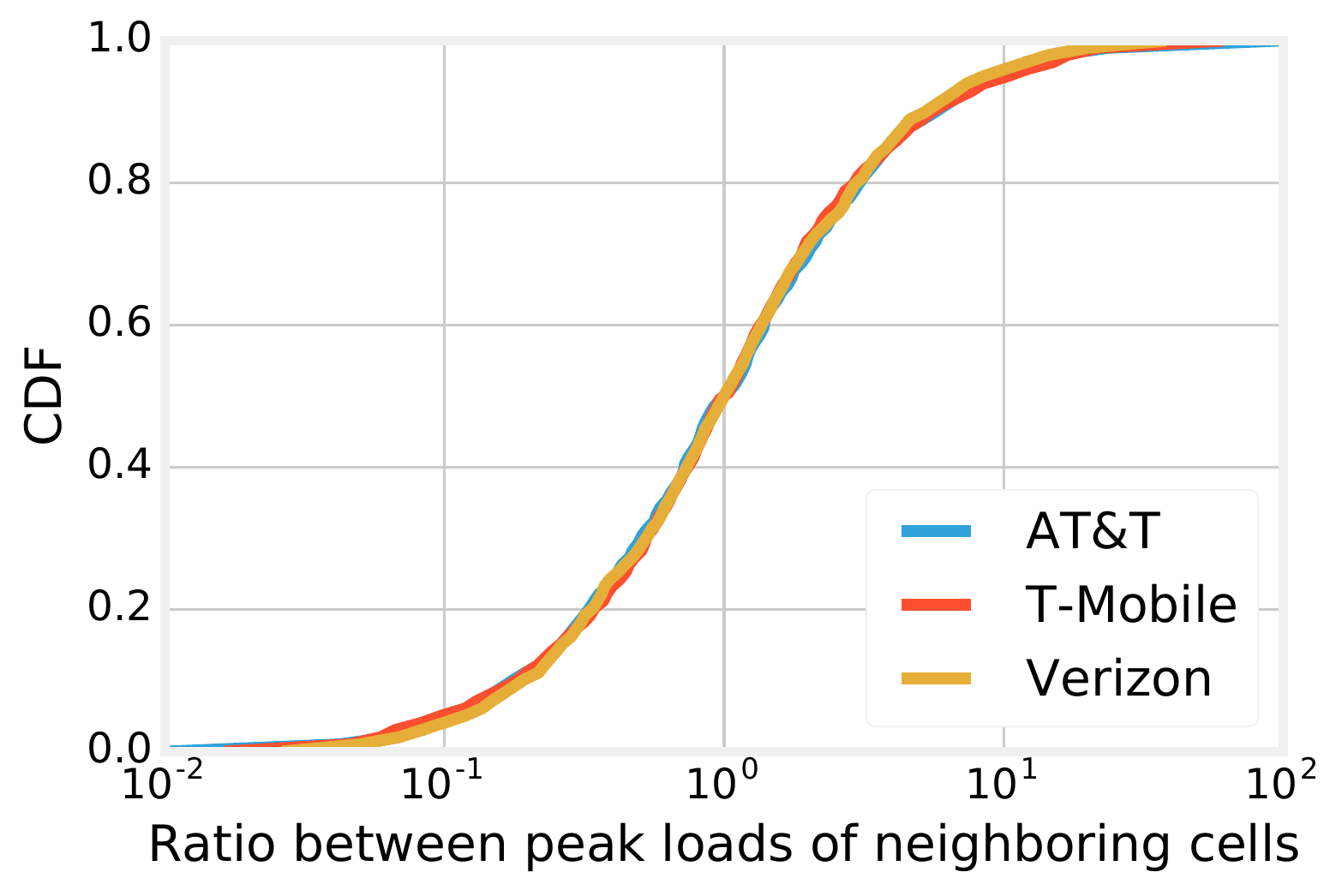}
}
\caption{
How MEC can improve server utilization: toy example (a); distribution of the peak load across cells in San Francisco (b); distribution of the ratio between peak loads of neighboring cells (c).
\label{fig:explain}
} 
\end{figure*}

\begin{figure*}[h!]
\centering
\subfigure[\label{fig:traffics-atl}]{
    \includegraphics[width=.3\textwidth]{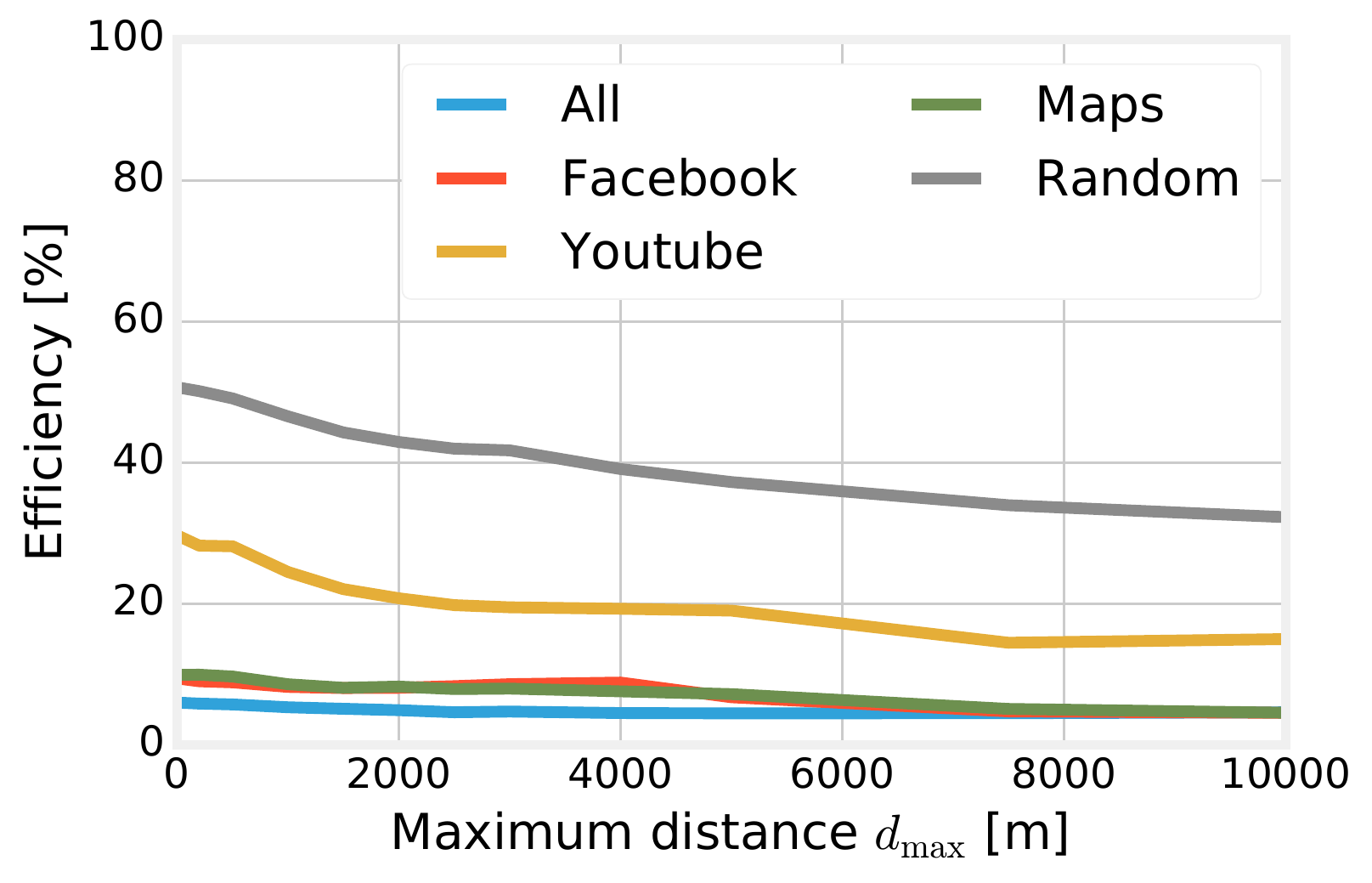}
}
\subfigure[\label{fig:traffics-la}]{
    \includegraphics[width=.3\textwidth]{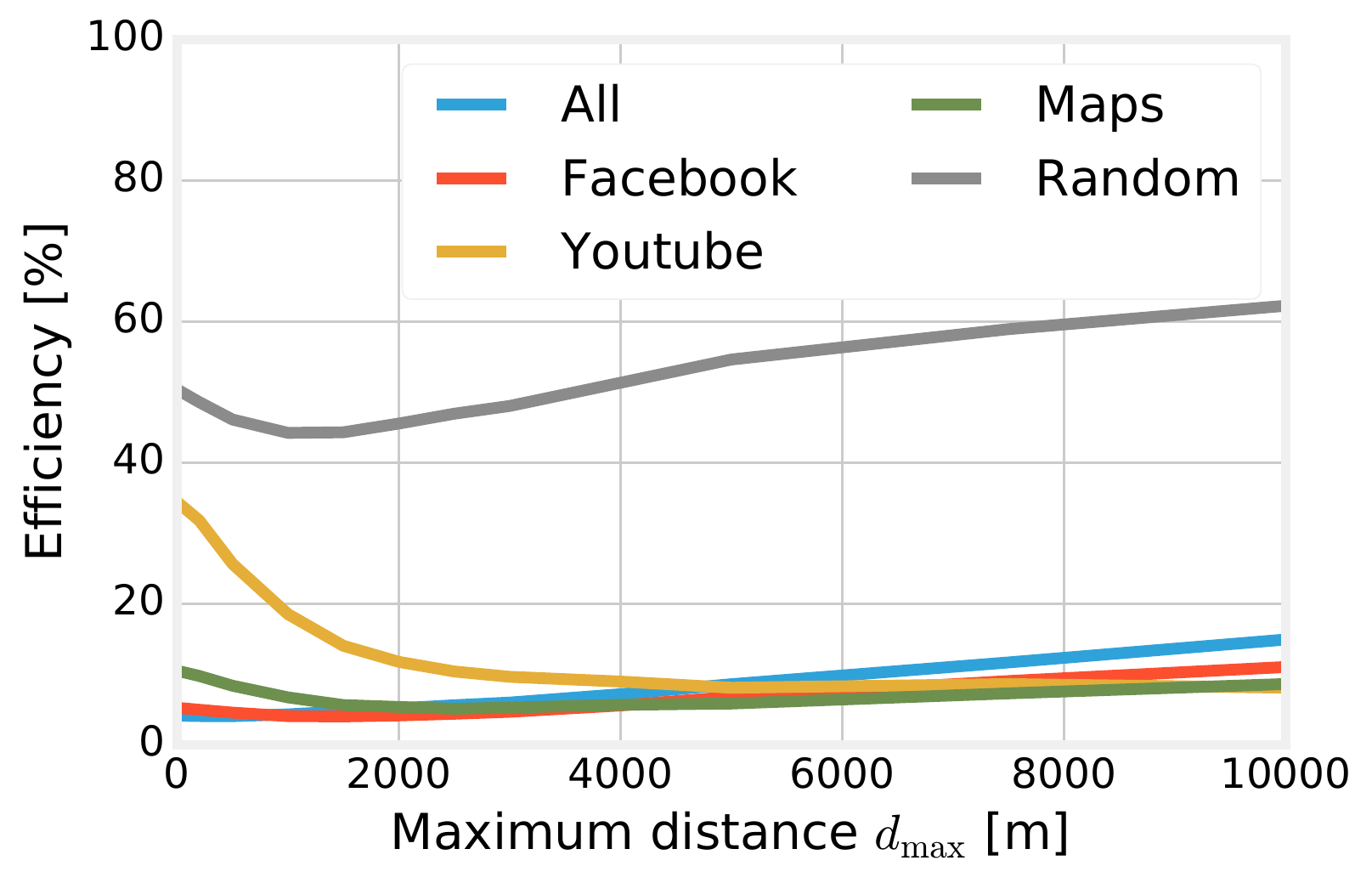}
}
\subfigure[\label{fig:traffics-sf}]{
    \includegraphics[width=.3\textwidth]{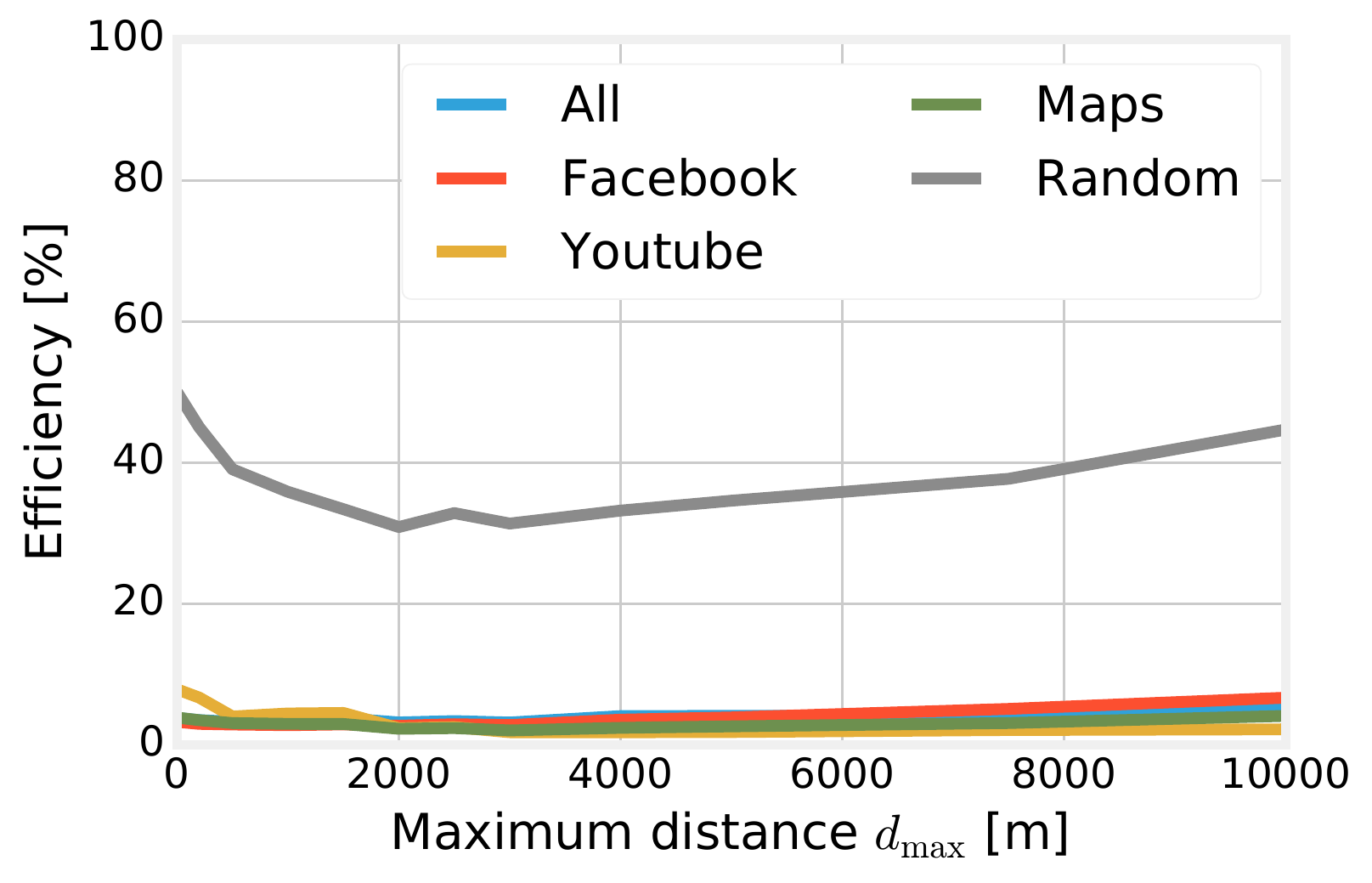}
}
\caption{
Evolution of our efficiency metric as a function of the maximum distance for different types of traffic for T-Mobile in Atlanta (a); Los Angeles (b); San Francisco (c).
\label{fig:traffics}
} 
\end{figure*}

\subsection{Server utilization}
\label{sec:results-efficiency}

We now move to studying server utilization, through our efficiency proxy metric: \Fig{efficiency} studies how the average
efficiency changes throughout cities and mobile operators. Consistently with the well-known fact that present-day cellular networks are overprovisioned~\cite{5gcloud,Malandrino-TCS}, we can observe that server utilization is always quite low, with an efficiency lower than~50\% even when the clusters have become fairly large.
The other insight we obtain from \Fig{efficiency} is quite surprising: the highest efficiency values are attained for the {\em low} values of~$d_{\max}$, not higher ones. This contrasts with our expectation and intuition
and seems to suggest that embracing the MEC paradigm allows us to reduce the latency {\em and} improve server utilization.

While this is undeniably good news, understanding the reason for this effect deserves some more careful analysis.
The reason for the discrepancy between our intuition and the results shown in \Fig{efficiency} lies in an implicit assumption we made, i.e., that the {\em peak} loads of different base stations were similar or at least comparable.

Consider, instead, the case of \Fig{example2}, where the yellow base station has a much higher peak load than the others, yet each has a similar average load. If we serve all three base stations with the same server, such a server would have a 15\% utilization. Serving that load with separate servers, instead, yields a combined utilization of 21\%. In other words, if the peak load of base stations being served by the same server is too different, then server utilization might decrease as a result.

\Fig{example2} suggests one possible explanation for the results we observe in \Fig{efficiency}; we now have to verify whether that is indeed the case using our real-world traces. \Fig{loadcdf} shows the distribution of the peak load for different base stations in San Francisco (results for other cities are omitted for brevity). We can observe that the $x$-scale in \Fig{loadcdf} spans multiple orders of magnitude: base stations whose peak loads are as different as the ones in \Fig{example2} (and indeed, much more) are very common in our traces.
However, we still do not know if base stations with wildly different peak loads tend to be close to each other, i.e., to end up in the same cluster. To this end, we consider all pairs of {\em neighboring} cells, and study the ratio between their peak loads. This allows us to determine whether cells served by the same server have similar peak loads
or very different ones (as in \Fig{example2}). In \Fig{ratios}, we take into account all pairs of neighboring base stations (i.e., base stations whose coverage areas overlap), and plot the distribution of the ratios between their peak loads: we can clearly see that, given a cell, we have a 60\% probability that it will have at least one neighbor whose peak load is more than two orders of magnitude apart, i.e., 100~times higher or lower. Apparently, the situation exemplified in \Fig{example2} is very common.

It is important to remark that, when the distance~$d_{\max}$ becomes very high, we do indeed observe an improvement in the server utilization, as reported in \Fig{efficiency}. This is the effect we were expecting earlier, the one
we were expecting: that combining base stations would smooth workload fluctuations and improve server utilization.

\subsection{Applications}
\label{sec:results-applications}

Different types of traffic have different latency requirements, and therefore the same latency and utilization that would be acceptable for an application (e.g., Internet-of-things) and unacceptable for another (e.g., gaming). For this reason, content providers are often envisioned~\cite{Malandrino-TCS} to deploy their own core and/or access infrastructure in oder to serve their own users.
This leads to the question of whether some types of traffic are more suitable than others to being processed in a MEC fashion. Thanks to the fact that our real-world traces include information on the actual applications active on the users' devices, we can  answer such a question.

\Fig{traffics} summarizes how server utilization changes across types of traffic and cities (plots are for T-Mobile; other operators have a similar behavior). A first aspect we notice is that there are significant differences between cities, a further reminder that MEC decisions ought to be made on a case-by-case basis. We also notice that YouTube traffic seems to consistently perform slightly better than the others. Keeping \Fig{explain} in mind, we can recall that YouTube sessions tend to be more homogeneous in size than other types of applications, as they all involve playing videos. This is especially relevant as video services are considered to be among those in direst need of MEC-style processing.

It is also important to point out the ``random'' curves, where real-world traffic is replaced by random values, selected with uniform probability between~$0$ and the maximum value we observe in the traces: they always correspond to a much higher value of the efficiency metric than real traffic. Indeed, assigning traffic at random makes it possible to attain all the gains
we would intuitively expect,
while at the same time minimizing the likelihood of effects such as the one in \Fig{example2} -- which results in an unrealistically optimistic estimation of server utilization. This serves as a further reminder of the importance of real-world traces as a tool to study real-world networks.

\section{Conclusion and future work}
\label{sec:conclusion}

We address a MEC scenario, where cloud servers are moved from the Internet to mobile networks themselves, close to the users whose traffic they process. We studied
the relationship
between reduced latency -- one of the main advantages of embracing the MEC paradigm -- and lower server utilization -- one of its potential drawbacks. To this end, we employed a set of large-scale, real-world, crowd-sourced traces, described in \Sec{traces}. As detailed in \Sec{steps}, we carried our study out using two proxy metrics that can be extracted directly from the traces themselves, allowing us to dispense with any further assumption.

Our results (\Sec{results}) show that MEC is able to yield both a short latency {\em and} a very high server utilization. We found the main reason for this to be the very different peak loads of different cells, including neighboring ones, whose traffic is likely to be processed by the same server.

We plan to extend our work by replacing the Voorhees algorithm with a custom-made clustering strategy,
accounting for traffic demand and server utilization.

\section*{Acknowledgement}
This work has received funding from the 5G-Crosshaul project (H2020-671598), and from the German-Israeli Cyber Security Center at the Hebrew University in Jerusalem.

\bibliographystyle{IEEEtran}
\bibliography{refs}

\begin{thebibliography}{10}
\providecommand{\url}[1]{#1}
\csname url@samestyle\endcsname
\providecommand{\newblock}{\relax}
\providecommand{\bibinfo}[2]{#2}
\providecommand{\BIBentrySTDinterwordspacing}{\spaceskip=0pt\relax}
\providecommand{\BIBentryALTinterwordstretchfactor}{4}
\providecommand{\BIBentryALTinterwordspacing}{\spaceskip=\fontdimen2\font plus
\BIBentryALTinterwordstretchfactor\fontdimen3\font minus
  \fontdimen4\font\relax}
\providecommand{\BIBforeignlanguage}[2]{{%
\expandafter\ifx\csname l@#1\endcsname\relax
\typeout{** WARNING: IEEEtran.bst: No hyphenation pattern has been}%
\typeout{** loaded for the language `#1'. Using the pattern for}%
\typeout{** the default language instead.}%
\else
\language=\csname l@#1\endcsname
\fi
#2}}
\providecommand{\BIBdecl}{\relax}
\BIBdecl

\bibitem{noi-pof}
F.~Malandrino, C.~Chiasserini, and S.~Kirkpatrick, ``The price of fog: A
  data-driven study on caching architectures in vehicular networks,'' in
  \emph{ACM MobiHoc IoV-VoI Workshop}, 2016.

\bibitem{orange-d4d}
V.~D. Blondel, M.~Esch, C.~Chan, F.~Cl{\'e}rot, P.~Deville, E.~Huens,
  F.~Morlot, Z.~Smoreda, and C.~Ziemlicki, ``Data for development: the d4d
  challenge on mobile phone data,'' \emph{arXiv preprint}, 2012.

\bibitem{Malandrino-TCS}
P.~D. Francesco, F.~Malandrino, T.~K. Forde, and L.~A. DaSilva, ``A sharing-
  and competition-aware framework for cellular network evolution planning,''
  \emph{IEEE Trans. on Cognitive Comm. and Netw.}, 2015.

\bibitem{wefi}
``{WeFi},'' \url{http://www.wefi.com}.

\bibitem{mit-reality}
N.~Eagle and A.~Pentland, ``Reality mining: sensing complex social systems,''
  \emph{Personal and ubiquitous computing}, 2006.

\bibitem{nokia}
J.~K. Laurila, D.~Gatica-Perez, I.~Aad, O.~Bornet, T.-M.-T. Do, O.~Dousse,
  J.~Eberle, M.~Miettinen \emph{et~al.}, ``The mobile data challenge: Big data
  for mobile computing research,'' in \emph{Pervasive Computing}, 2012.

\bibitem{fog}
Cisco, ``{Transform Data into Action at the Network Edge},'' 2015.

\bibitem{etsi-wp}
{ETSI}. Mobile edge computing white ppaer.
  \url{http://www.etsi.org/technologies-clusters/technologies/mobile-edge-computing}.

\bibitem{mec-radio}
S.~Sardellitti, G.~Scutari, and S.~Barbarossa, ``Joint optimization of radio
  and computational resources for multicell mobile-edge computing,'' \emph{IEEE
  Transactions on Signal and Information Processing over Networks}, 2015.

\bibitem{mec-iot}
F.~Bonomi, R.~Milito, J.~Zhu, and S.~Addepalli, ``Fog computing and its role in
  the internet of things,'' in \emph{Proceedings of the First Edition of the
  MCC Workshop on Mobile Cloud Computing}, 2012.

\bibitem{mec-5g}
S.~Nunna, A.~Kousaridas, M.~Ibrahim, M.~Dillinger, C.~Thuemmler, H.~Feussner,
  and A.~Schneider, ``Enabling real-time context-aware collaboration through 5g
  and mobile edge computing,'' in \emph{ITNG}, 2015.

\bibitem{moving}
F.~Sardis, G.~Mapp, J.~Loo, and M.~Aiash, ``Dynamic edge-caching for mobile
  users: Minimising inter-as traffic by moving cloud services and vms,'' in
  \emph{IEEE WAINA}, 2014.

\bibitem{multiop-caching}
K.~Poularakis, G.~Iosifidis, A.~Argyriou, I.~Koutsopoulos, and L.~Tassiulas,
  ``Caching and operator cooperation policies for layered video content
  delivery,'' in \emph{IEEE INFOCOM}, 2016.

\bibitem{cdn1}
X.~Cai, S.~Zhang, and Y.~Zhang, ``Economic analysis of cache location in mobile
  network,'' in \emph{IEEE WCNC}, 2013.

\bibitem{proactive-caching}
E.~Bastug, M.~Bennis, and M.~Debbah, ``{Living on the edge: The role of
  proactive caching in 5G wireless networks},'' \emph{IEEE Comm. Mag.}, 2014.

\bibitem{wons}
``Mobility-aware edge caching for connected cars,'' in \emph{WONS}, 2016.

\bibitem{softcell}
X.~Jin, L.~E. Li, L.~Vanbever, and J.~Rexford, ``{Softcell: Scalable and
  flexible cellular core network architecture},'' in \emph{{ACM CoNEXT}}, 2013.

\bibitem{clustering}
E.~M. Voorhees, ``Implementing agglomerative hierarchic clustering algorithms
  for use in document retrieval,'' \emph{Information Processing \& Management},
  1986.

\bibitem{5gcloud}
K.~Zheng, T.~Taleb, A.~Ksentini, C.~L. I, T.~Magedanz, and M.~Ulema, ``Research
  and standards: advanced cloud and virtualization techniques for 5g networks
  (part ii) [guest editorial],'' \emph{IEEE Communications Magazine}, 2015.

\end{thebibliography}
\balancecolumns 

\end{document}